# Tuning up the performance of GaAs-based solar cells by inelastic scattering on quantum dots and doping of $Al_yGa_{1-y}Sb$ type-II dots and $Al_xGa_{1-x}As$ spacers between dots

A. Kechiantz*, A. Afanasev
Department of Physics, The George Washington Univ., 725 21st Street, NW, Washington, DC, USA, 20052, kechiantz@gwu.edu , afanas@gwu.edu ; *On leave: Institute of Radiophysics and Electronics, National Academy of Sciences, 1 Brothers Alikhanyan st., Ashtarak 0203, Armenia

## ABSTRACT

We used AlGaSb/AlGaAs material system for a theoretical study of photovoltaic performance of the proposed GaAs-based solar cell in which the type-II quantum dot (QDs) absorber is spatially separated from the depletion region. Due to inelastic scattering of photoelectrons on QDs and proper doping of both QDs and their spacers, concentrated sunlight is predicted to quench recombination through QDs. Our calculation shows that 500-sun concentration can increase the Shockley-Queisser limit from 35% to 40% for GaAs single-junction solar cells.

## 1 INTRODUCTION

The Shockley-Queisser limit for conversion efficiency of solar cells is calculated in the framework of the principle of detailed balance assuming ideal conditions of only radiative inter-band electron transitions in the cell[1]. The model assumes that photoelectrons generated by the above-band gap photons swiftly relax to the conduction band edge by transferring their excess energy to the semiconductor lattice due to the intra-band scattering on optical phonons. This relaxation absorbs about 30% of solar energy. Our goal is to put into use this excess energy of photoelectrons.

The sub-band gap photons compose another 30% of solar energy that the Shockley-Queisser model of ideal p-n-junction misses. However, the intermediate band (IB) concept by Luque and Marti suggests a way for putting into use the energy of sub-band gap photons[2], in fact, due to the non-linear effect of two-photon absorption[3]. This concept assumes that consecutive absorption of two sub-band gap photons, whose combined energy exceeds the semiconductor band-gap, may generate an additional photocurrent by transferring an electron from the valence band into the conduction band. The IB concept involves a band of intermediate electronic states resonant with sub-band gap photons within the semiconductor band gap for facilitating their two-photon absorption[4].

The IB absorber can be composed with quantum dots (QDs)[5], which are zero-dimensional semiconductor building blocks with unique electronic properties. While Luque and Marti proposed to use the two-photon absorption in QDs for generating of additional photocurrent in IB solar cells[5], Petrosyan with co-authors argued that simply combining electron tunneling from QDs with conventional single-photon absorption of sub-band gap photons in QDs could generate the additional photocurrent[6]. In both cases type-I QDs were located within the depletion region of p-n-junction. Both such location and type-I QDs facilitate recombination processes through QDs[7], which increases the dark current and hence reduces the open circuit voltage of solar cells[2,3].

Spatial separation of the QD absorber from the depletion region is an evident solution of this problem since the depletion region is the most sensitive part of solar cells where electronic states easily facilitate recombination processes[3,7]. If the QD absorber would be placed within the photoelectron diffusion length from the depletion region, the p-n-junction would still collect all photoelectrons. The next step towards extinguishing the additional dark current generated through QD is substitution of type-I for type-II QDs[7]. While type-I QDs easily capture photoelectrons from the conduction band and holes from the valence band for recombination, type-II QDs spatially separate mobile electrons of conduction band from holes confined in QDs. Such separation decreases recombination through QDs. For instance, in GaSb/GaAs strained semiconductor systems lifetime of mobile electrons with respect to recombination with holes confined in GaSb type-II QDs is very close to the minority carrier lifetime in GaAs, $10 ns$[8,9]. Our calculations shown that the separation of type-II QD absorber from the depletion region eliminates the additional dark current[3,7,10].

The spatial separation of QD absorber also reduces the photocurrent for the following two reasons. First, the holes generated by the above-band gap photons accumulate in QDs; second, inter-band absorption of sub-band gap photons weakens when too many holes are confined in QDs. We have already shown that concentration of sunlight facilitates the two-photon absorption of sub-band gap photons in separated type-II QD absorber so much that the conversion efficiency achieves the Luque-Marti limit when the concentration becomes about 300-suns[11].

In this paper we emphasize that inelastic scattering of photoelectrons on QDs may eliminate recombination through QDs by swiftly removing confined holes from QDs. Our main idea is that if there are no holes in confined states, photoelectrons have to miss transitions from the conduction band into such QD confined electronic states. Such QDs will absorb sub-band gap photons and escape recombination processes, hence, will not increase the dark current. These QDs will facilitate the additional photocurrent if we swiftly remove photo-generated holes from QDs into mobile electronic states in the valence band. While the IB concept exploits the next sub-band gap photons for removing photo-generated holes from QDs, we propose to involve inelastic scattering of photoelectrons on QDs for the same removing of confined holes.

Though QD confined electronic states are discrete, they have a high local density of confined states in the semiconductor valence band and are spread from the valence band edge deep into the fundamental band gap[12,13]. We show that such distribution of confined states may change the intra-band relaxation mechanism that photoelectrons transfer their excess energy to the semiconductor lattice. Our calculation shows that the intra-band relaxation may dominate by inelastic scattering on QDs and facilitate the escape of confined holes from QDs. The latter depends on both doping of QD absorber and concentration of sunlight. While such escape has no influence on the dark current, it increases generation of additional photocurrent by one-photon absorption of concentrated sunlight. We refer to the GaSb/GaAs strained semiconductor system in our calculation because the large offset, type-II (staggered) misalignment of the conduction and valence bands, direct band gaps, and well-developed fabrication technology make this system a good object for our study.

## 2   QD ABSORBER SEPARATED FROM THE DEPLETION REGION

### 2.1  Design

In the previous paper[11] we have described the design and operation principle of IB solar cell with spatially separated type-II QD absorber of sub-band gap photons, as shown in Figure 1.

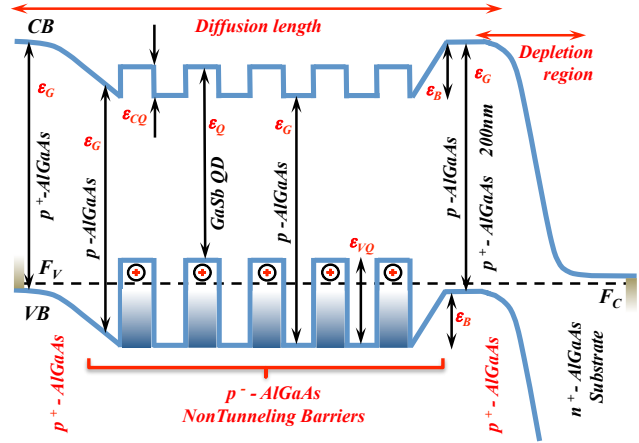

Figure 1. Energy band diagram of IB solar cell with a spatially separated GaSb/GaAs type-II QD absorber[11]. The stack of GaSb QDs is embedded in $p^+$-doped $Al_xGa_{1-x}As$ layer so that $p^+$-doped $Al_xGa_{1-x}As$ layer separates the stack from the depletion region. The $Al_xGa_{1-x}As$ barriers in valence band are non-tunneling for holes confined in QDs.

This absorber is an epitaxial stack comprising GaSb strained QD layers alternating with p-doped $Al_xGa_{1-x}As$ spacers such that spacers compose non-tunneling barrier layers around QDs preventing escape of confined holes from QDs in the valence band. The absorber is sandwiched between a $p^+$-doped $Al_xGa_{1-x}As$ cap layer and a thin $p^+$-doped $Al_xGa_{1-x}As$ buffer layer follows an $n^+$-doped $Al_xGa_{1-x}As$ buffer layer grown on an $n^+$-doped GaAs substrate. It is assumed that buffers compose an ideal $Al_xGa_{1-x}As$ p-n-junction. The absorber is spatially separated from the depletion region of the $Al_xGa_{1-x}As$ p-n-junction is located within the electron diffusion length distance from that region. Therefore, all photoelectrons generated in the QD absorber diffuse towards the p-n-junction and contribute into photocurrent. At the same time, electron tunneling through the depletion region into the electronic states confined in QDs becomes impossible; hence, QDs cannot facilitate generation of additional dark current reducing the open circuit voltage of IB solar cells. An important feature of this design is that the QD absorber is doped less than the p$^+$-doped Al$_x$Ga$_{1-x}$As cap and buffer layers. For equalizing Fermi level across the cell such doping introduces holes from both cap and buffer layers into QDs, which keeps the confined states above the Fermi level as shown in Figure 1.

Let us focus on the QD absorber that keeps the confined states below the Fermi level, as shown in Figure 2. Due to such specific doping profile, there are only a few holes either mobile or confined in QD absorber. Therefore, the absorber is about transparent for $\hbar\omega_2$ sub-band gap photons from $\varepsilon_{VQ} < \hbar\omega_2 < \varepsilon_Q$ spectral range. At the same time the one-photon inter-band absorption of both $\hbar\omega_1$ sub-band gap and $\hbar\omega$ above-band gap photons from $\varepsilon_Q < \hbar\omega_1 < \varepsilon_G < \hbar\omega$ spectral ranges is very strong in such QD absorber. Also the blocking barrier $\varepsilon_B$ is higher than that in QD absorber shown in Figure 1. The QD absorber is about 1 $\mu m$-thick such that it absorbs all incoming photons from $\varepsilon_Q < \hbar\omega$ spectral range. The p$^+$-doped Al$_x$Ga$_{1-x}$As cap layer is thin to be transparent for the above-band gap energy photons. Aluminum content, $x$, of Al$_x$Ga$_{1-x}$As in both buffer p-n-junction and cap layer is higher than it is in Al$_x$Ga$_{1-x}$As spacers of the QD absorber so that $\varepsilon_G < \varepsilon_{BF} \leq \varepsilon_{CAP}$. The p$^+$-doped wide band gap optical window caps the structure.

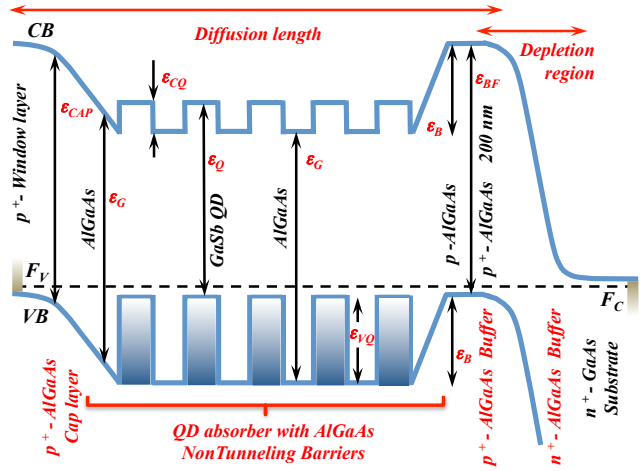

Figure 2. Energy band diagram of solar cell with separated GaSb/GaAs type-II QD absorber that the confined states are below the Fermi level. The Al$_x$Ga$_{1-x}$As barriers in valence band are non-tunneling for holes confined in QDs.

## 2.2 Recombination through QDs

The spatial separation of mobile electrons from confined holes in type-II QDs increases the lifetime associated with inter-band recombination through QDs[9]. The pump-probe study of photoluminescence from GaSb/GaAs strained type-II QDs demonstrated 10 $ns$ decay time[8]. However, those QDs were overcrowded with confined holes while QDs in proposed absorber shown in Figure 2 have very few holes. Therefore, the photoelectron lifetime in conduction band of proposed absorber is expected to increase reversal to density of confined holes in absorber and to be longer than 10 $ns$.

Absorption of concentrated sunlight rearranges distribution of charge carriers in QD absorber, reduces the blocking barrier $\varepsilon_B$, and splits the Fermi level into quasi-Fermi levels for mobile electrons in conduction band, mobile holes in valence band, and confined holes in QDs. However, the rearrangement does not reduce the photoelectron lifetime in conduction band of QD absorber until there are very few holes confined in QDs. Therefore, a mechanism should be involved for swift removal of photo-generated holes from QDs and keeping the confined states below their quasi-Fermi level as shown in Figure 3.

Figure 3 displays how two consecutive absorptions of sub-band gap photons generate a confined hole and remove it from a QD. The red dashed arrows in Figure 3 denote the electron transfers to higher energy states due to photon absorption. While absorption of $\hbar\omega_1$ sub-band gap photons, $\varepsilon_Q < \hbar\omega_1 < \varepsilon_G$, generates confined holes in QDs by removing electrons from QDs into the conduction band, the absorption of $\hbar\omega_2$ sub-band gap photons, $\varepsilon_{VQ} < \hbar\omega_2 < \varepsilon_Q$, eliminates those confined holes from QDs by removing them into the mobile electronic states of the valence band in absorber.

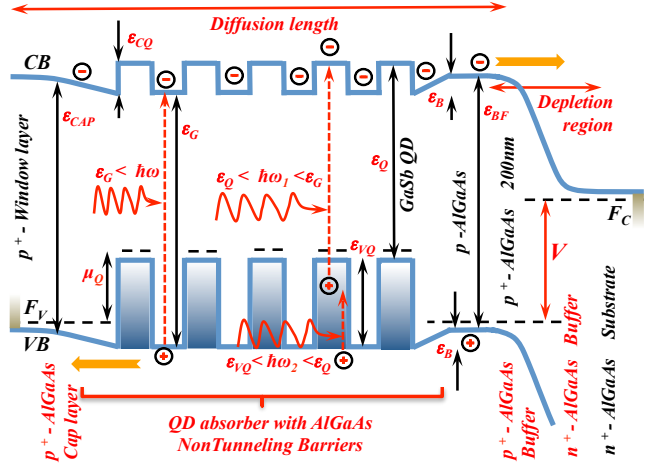

Figure 3. Energy band diagram of GaSb/GaAs type-II QD IB solar cell under concentrated sunlight illumination. The negative charge of photoelectrons accumulated in the conduction band of QD absorber and the positive charge accumulated in the buffer move up the conduction band edge reducing the blocking barrier $\varepsilon_B$ as compared to that shown in Figure 2.

## 2.3 Photoelectron relaxation

Just generated photoelectrons have excess energy that they release in picoseconds due to the intra-band scattering on optical phonons. For a $ps$ time scale the photoelectron diffuses $100nm$ long distance and a few times scatters on QDs since the average distance between QDs is about $30nm$. If the excess energy is above $\varepsilon_{VQ}$ and a hole is confined in the QD as shown in Figure 4, the photoelectron can release the hole from the QD by transferring its excess energy to the hole. Such relaxation of photoelectron excess energy can be more effective than relaxation on optical phonons if the events occur often because the energy release per event is larger that the energy of optical phonons by an order of magnitude.

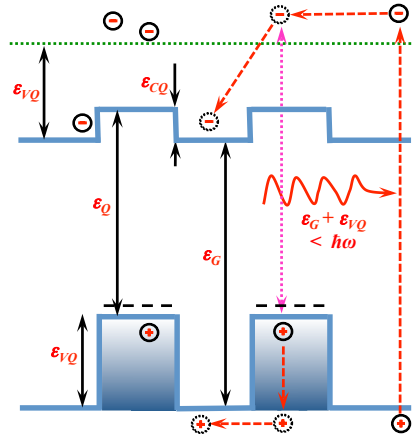

Figure 4. Energy relaxation due to inelastic scattering of photoelectrons on QDs. Above the green-dot line, photoelectrons have enough energy for removing confined holes from QDs.

The intensity $I_{is}$ of inelastic scattering per QD can be written as $I_{is} = X\,G_{he}(v_T \tau_{oph})^3/w_a$. Assuming absorption of $X = 1000$ times concentrated sunlight; $w_a = 1\mu m$ thick QD absorber; $\tau_{oph} = 1ps$ electron-phonon relaxation time; $v_T = 10^7\ cm/s$ electron thermal velocity; and $G_{he} = 10^{17} cm^{-2} s^{-1}$ intensity of $\hbar\omega_{he}$ high energy photons in solar spectrum, $\varepsilon_Q + \varepsilon_{VQ} < \hbar\omega_{he}$, we can calculate $I_{is} = 10^9 s^{-1}$ so that within $1ns$ the confined holes acquire energy to escape from QDs into the mobile electronic states of the valence band in absorber. The intensity $I_{\omega 1}$ of generation of confined holes per QD by sub-band gap photons of the same concentrated sunlight can be written as $I_{\omega 1} = X\,G_{\omega 1}\Omega/w_a$.

Assuming $G_{\omega 1} = 10^{17} cm^{-2} s^{-1}$ intensity of $\hbar\omega_1$ sub-band gap photons in solar spectrum, $\varepsilon_Q < \hbar\omega_1 < \varepsilon_G$, and $\Omega = 10^{-18} cm^3$ volume per QD, we can calculate $I_{\omega 1} = 10^6 s^{-1}$. This reduces the scattering per hole-generation in a QD, $I_{is}/I_{\omega 1} = (G_{he}/G_{\omega 1})(v_T \tau_{oph})^3/\Omega$, to $I_{is}/I_{\omega 1} = 10^3$ so that the next confined hole will be generated in a QD after thousand photoelectron-QD scattering events only. Thus one can see that inelastic scattering of high-energy photons becomes the dominant mechanism for removing of confined holes from QDs into the mobile electronic states of the valence band in absorber.

### 2.4 Balance of electron transitions through QDs

Absorption of concentrated sunlight and bias voltage bring the quasi-Fermi level of QD confined states to the position that makes current, $j$, from QDs into the conduction band equal to the current from the valence band into QDs,

$$j = j_{SVC} - j_{RCV}[exp(eV/kT) - 1] + j_{SQC} - j_{RQC}\left(exp\frac{eV-\mu}{kT} - 1\right) \tag{1}$$

$$j_{SQC} - j_{RQC}\left(exp\frac{eV-\mu}{kT} - 1\right) = (j_{SVQ} - j_{RVQ}\left(exp\frac{\mu}{kT} - 1\right) + j_{IS}\frac{(\tau_{ph} v_T)^3}{\Omega})exp\frac{\varepsilon_{QV}-\mu+\varepsilon_B}{kT} - j_{PI} \tag{2}$$

$$j_{ji} = \frac{2eX}{h^3 c^2 GF}\int_{\varepsilon_i}^{\varepsilon_j}\frac{exp[(-\mu)/kT]\varepsilon^2 d\varepsilon}{exp[(\varepsilon-\mu)/kT]-1} \tag{3}$$

where in case of photocurrents related to solar photon absorption, $X$ is the concentration of solar light, $\mu = 0$, and $GF = 4.6\times10^4$ is the geometrical factor related to the angle that Earth is seen from Sun; in case of radiative recombination currents related to photon emission from the solar cell, $X = GF = 1$ and $\mu$ is the splitting of quasi-Fermi levels; $[\varepsilon_i, \varepsilon_j]$ are the spectral ranges, $i$ and $j$ refer to either the conduction band, $C$, or the valence band, $V$, or QDs, $Q$; $R$ denotes recombination, $j_{IS}$ is the current associated to the inelastic scattering of photoelectrons on QDs, $j_{PI}$ is the injection current of holes into absorber from the $p^+$-doped $Al_xGa_{1-x}As$ cap layer, $j_{PI} = (ep_0 w_a/\tau_{ph})exp(-\varepsilon_B/kT)$; $\tau_{ph} = 1 ps$ is the electron-phonon relaxation time; $v_T$ is the thermal velocity of photoelectrons; $\Omega = 10^{-18} cm^3$ is the volume per QD; $w_a = 1\mu m$ is the thickness of the QD absorber; $p_0$ is the concentration of holes in the cap and buffer $p^+$-doped $Al_xGa_{1-x}As$ layers; we used the principle of detailed balance to reduce currents generated by radiative electron transitions to integrals[2]. We assume absorption of all incoming solar photons that $\varepsilon_Q < \hbar\omega$. The coefficient $exp[-(\mu + \varepsilon_B - \varepsilon_{QV})/kT]$ appears in Equation 2 because the effectiveness of currents associated with photoelectron inelastic scattering on QDs or sub-band gap photon absorption in QDs is proportional to the density of confined holes; $\mu = \mu_Q$ is the quasi-Fermi level for QD confined states as shown in Figure 3; $kT$ is the temperature of the cell; $V$ is the bias voltage; $eV - \mu$ is the separation of the conduction band quasi-Fermi level from that of QDs; and $\varepsilon_B$ is the blocking barrier that limits the flow of holes into the QD absorber from the cap and buffer layers. The charge accumulated in the conduction band in QD absorber determines the height $\varepsilon_B$ of this blocking barrier.

## 3  RESULTS AND DISCUSSION

### 3.1 Accumulation of photoelectrons in QD absorber

Solution of Equations (1) - (3) enables us to describe the photovoltaic characteristics of the proposed GaAs-based solar cell with GaSb type-II QD absorber spatially separated from the depletion region. The doping profile lowers the conduction and valence band edges of absorber relative to that in $p^+$-doped $Al_xGa_{1-x}As$ cap and buffer layers by $\varepsilon_B$ as shown in Figure 2. This reduction does not reach the depletion region of p-n-junction, therefore photo-generated electrons face a blocking barrier $\varepsilon_B$ in the conduction band on their way to the p-n-junction and suaccumulate in absorber. The same barrier blocks injection of mobile holes into absorber from the $p^+$-doped $Al_xGa_{1-x}As$ cap and buffer layers. When confined holes hop from QDs into the mobile electronic states of the valence band they swiftly escape QDs and absorber for the $p^+$-doped $Al_xGa_{1-x}As$ cap layer. The positive charge of holes accumulated in the $p^+$-doped $Al_xGa_{1-x}As$ buffer layer balances the diffusion of generated holes into the buffer.

By the same time the negative charge of photoelectrons accumulated in the absorber raise the conduction and valence band edges. This reduces the blocking barrier $\varepsilon_B$ as shown in Figure 3. The reduced barrier enables injection of mobile holes from the p$^+$-doped $Al_xGa_{1-x}As$ cap layer into absorber. These holes bring a positive charge that limits the reduction of blocking barrier $\varepsilon_B$. Therefore, emission of photo-generated holes, $X(G_{\omega 1} + G_\omega)$, has to be more intensive that the injection of mobile holes, $(p_{cap}D_h/w_a)exp(-\varepsilon_B/kT)$, which gives $(p_{cap}D_h/w_a)exp(-\varepsilon_B/kT) \leq X(G_{\omega 1} + G_\omega)$. Assuming $p_{cap} = 10^{18}\ cm^{-3}$ density of holes in the p$^+$-doped $Al_xGa_{1-x}As$ cap layer; $D_h = 10\ cm^2 s^{-1}$ the diffusion coefficient of holes; and $G_\omega = 10^{17} cm^{-2}s^{-1}$ the intensity of $\hbar\omega$ above-band gap photons in the solar spectrum, $\varepsilon_G < \hbar\omega$, we can find that concentration should be $X \geq 10^6 exp(-\varepsilon_B/kT)$. The Figure 5 displays this constraint. For instance, if $\varepsilon_B > 0.25eV$, already 100-sun concentration is enough for removing generating holes from QDs due to inelastic scattering of photoelectrons on QDs. However, low concentration limits the allowed bias voltage as shown in Figure 5. For 100-sun concentration, the allowed bias voltage is below $1eV$ while for 1000-sun concentration, the limit is $1.15eV$.

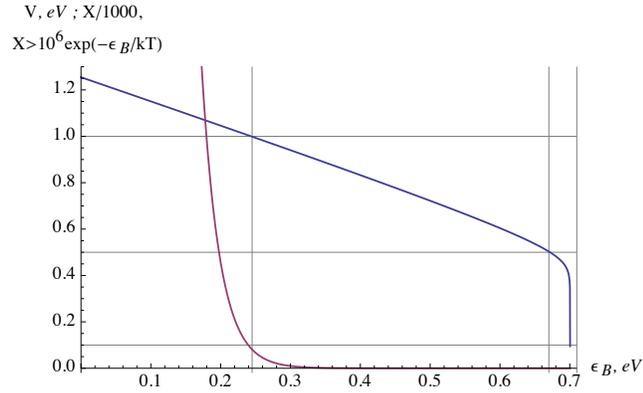

Figure 5. The constraint $X \geq 10^6 exp(-\varepsilon_B/kT)$ on the blocking barrier $\varepsilon_B$ and concentration of sunlight, (a); and the dependence of blocking barrier $\varepsilon_B$ on the bias voltage $V$, (b)..

Solution of Equations (1) –(3) makes it clear that these constraints on the blocking barrier are intimately connected with the position of quasi-Fermi level $\mu_Q$ of confined holes relative to the level of the QD confined ground electronic state. As shown in Figure 6, the quasi-Fermi level $\mu_Q$ is above the ground state by $30meV$ until $\varepsilon_B > \varepsilon_{QV} = 0.45eV$. Further reduction of the concentration or increase of the bias reduces the blocking barrier so much that the injection of holes from the cap layer becomes dominant for the inelastic scattering. As a result the quasi-Fermi level $\mu_Q$ enters into the energy region of confined electronic states.

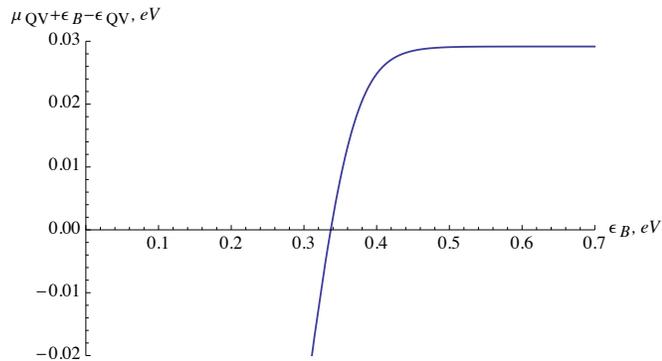

Figure 6. The quasi-Fermi level of confined holes $\mu_Q$ relative to the confined ground electronic state as a function of the blocking barrier $\varepsilon_B$.

## 3.2 Conversion efficiency

Reduction of the blocking barrier enables photoelectrons accumulated in the conduction band of QD absorber to pass through the depletion region of the p-n-junction into n-doped substrate. This way all electrons generated in conduction band of the QD absorber by incoming solar photons that $\varepsilon_Q < \hbar\omega$ contribute into the photocurrent. This photocurrent includes contribution of the sub-band gap photons that $\varepsilon_Q < \hbar\omega < \varepsilon_G$, therefore, the conversion efficiency $\eta$ of the proposed GaAs-based single-junction solar cell achieves 40% for 500-sun concentration as shown in Figure 7. This value is lower than the Luque-Marti limit of 50% for 500-sun concentration induced two-photon absorption of sub-band gap photons. However, for comparison, the same Figure 7 displays the Shockley-Queisser limit of GaAs solar cell, which is 35% for 500-sun concentration. Noteworthy, our proposed solar cell is not optimized yet.

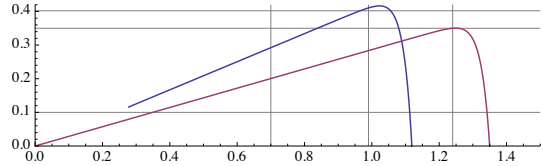

Figure 7. Conversion efficiency $\eta$ of GaAs-based solar cell with spatially separated GaSb type-II QD absorber as a function of bias voltage (blue line). Shockly-Queisser limit of GaAs solar cell (red line).

## 4 CONCLUSIONS

In conclusion, our study has shown that inelastic scattering on QDs can be a dominant mechanism of photoelectron intra-band relaxation in the proposed GaAs-based single-junction solar cell. Though electronic states confined in GaSb type-II QDs are discrete, they have a high local density of confined states in the valence band and are spread from the valence band edge deep into the fundamental band gap[12,13]. We demonstrated that such distribution of confined states has a potential to facilitate the photoelectron excess energy relaxation by increasing the photoelectron inelastic scattering on QDs. We have found conditions under which the energy transfer to the semiconductor lattice is dominated by the intra-band relaxation. We have shown that the inelastic scattering eliminates recombination through QDs due to the removal of confined holes from QDs so quickly that very few both confined and mobile holes stay in QD absorber. The inelastic scattering enables the sub-band gap photons to generate additional photocurrent, without involving QDs into generation of additional recombination current. Very importantly, the proposed cell design exploits type-II QDs spatially separated from the depletion region. These QDs cannot facilitate recombination in the QD absorber or current leakage through the depletion region. We refer to the GaSb/GaAs strained semiconductor system in our calculation because the large offset, type-II (staggered) misalignment of the conduction and valence bands, direct band gaps, and well-developed fabrication technology make this system a good object for our study. Our calculation has shown that the conversion efficiency limit of the proposed GaAs-based single-junction solar cell may reach 40% for 500-sun concentration, which is above the Shockley-Queisser limit of GaAs solar cell, 35%; however, this value is lower than the Luque-Marti limit of 50% for 500-sun concentration induced by two-photon absorption of sub-band gap photons, possibly because our cell is not optimized yet.

Recombination through QDs is a major factor that limits efficiency of solar cells based on QDs located within the depletion region. Our proposal will help to solve this problem.

## ACKNOWLEDGEMENTS

A. Kechiantz and A. Afanasev acknowledge support from The George Washington University.

## REFERENCES

[1] Shockley W. and Queisser H. J., "Detailed Balance Limit of Efficiency of p-n Junction Solar Cells", J. Appl. Phys. 32, **510-519 (1961)
[2] Luque A., Marty A., and Stanly C., "Understanding intermediate-band solar cells", Nature Photonics 6, 146-152,

(2012)

[3] Kechiantz A. M., Afanasev A., and Lazzari J.-L., "Efficiency limit of $Al_xGa_{1-x}As$ solar cell modified by $Al_yGa_{1-y}Sb$ quantum dot intermediate band embedded outside of the depletion region", Photovoltaic Tech. Conf.: Thin Film & Advanced Silicon Solutions, Aix-en-Provence, France (June 6-8, 2012)

[4] Luque A., Linares P. G., Antolın E., Ramiro I., Farmer C. D., Hernandez E., Tobıas I., Stanley C. R., and Martı A., "Understanding the operation of quantum dot intermediate band solar cells, J. Appl. Phys. 111, 044502 (2012)

[5] Luque A., and Marti A., "The Intermediate Band Solar Cell: Progress Toward the Realization of an Attractive Concept", Adv. Matter 22, 160-174 (2010)

[6] Aroutiounian V., Petrosyan S., Khachatryan A., and Touryan K., "Quantum dot solar cells", J. Appl. Phys. 89, 2268-2271 (2001)

[7] Kechiantz A.M., Kocharyan L.M., Kechiyants H.M., "Band alignment and conversion efficiency in Si/Ge type-II quantum dot intermediate band solar cells", Nanotechnology 18, 405401 (2007)

[8] Liang B., Lin A., Pavarelli N., Reyner C., Tatebayashi J., Nunna K., He J., Ochalski T.J., Huyet G., and Huffaker D.L., "GaSb/GaAs type-II quantum dots grown by droplet epitaxy", Nanotechnology 20, 455604 (2009)

[9] Sato D., Ota J., Nishikawa K., Takeda Y., Miyashita N., and Okada Y., "Extremely long carrier lifetime at intermediate states in wall-inserted type II quantum dot absorbers", J. Appl. Phys. 112, 094305 (2012)

[10] Kechiantz A. M., Afanasev A., Lazzari J.-L., Bhouri A., Cuminal Y., and Christol P., "Efficiency limit of $Al_xGa_{1-x}As$ solar cell modified by $Al_yGa_{1-y}Sb$ quantum dot intermediate band embedded outside of the depletion region", Proc. 27$^{th}$ EU PVSEC 16877, 412-417 (2012)

[11] Kechiantz A. M., Afanasev A., Lazzari J.-L., "Modification of band alignment at interface of $Al_yGa_{1-y}Sb/Al_xGa_{1-x}As$ type-II quantum dots by concentrated sunlight in intermediate band solar cells with separated absorption and depletion regions", SPIE Paper # 8620-20, SPIE Photonics West 2013 – OPTO, Optoelectronic Materials and Devices, Conference OE102: Physics, Simulation, and Photonic Engineering of Photovoltaic Devices II, The Moscone Center, San Francisco, California (USA), 2 - 7 February (2013)

[12] Geller M., Kapteyn C., Müller-Kirsch L., Heitz R., and Bimberg D., 450 meV hole localization in GaSb/GaAs quantum dots, Appl. Phys. Lett. 82, 2706-2708 (2003)

[13] Hwang J., Martin A. J., Millunchick J. M., and Phillips J. D., Thermal emission in type-II GaSb/GaAs quantum dots and prospects for intermediate band solar energy conversion, J. Appl. Phys. 111, 074514 (2012)